\documentclass{epl}
\input{epsf.tex}

\title{Probing the superconducting condensate on a nanometer scale}

\author{Th. Proslier, A. Kohen, Y. Noat, T. Cren, D. Roditchev and W. Sacks}

\shorttitle{}

\institute{Institut des Nano-Sciences de Paris, I.N.S.P.,
Universit\'es Paris 6 et Paris 7, et C.N.R.S. (UMR 75\ 88), 140
rue de Lourmel, Campus Boucicaut, 75015 Paris, France}

\pacs{74.50.+r}{First pacs description} \pacs{74.70.-b}{ Second
pacs description} \pacs{07.79.Cz}{ Third pacs description}

\begin{document}

\maketitle

\begin{abstract}

Superconductivity is a rare example of a quantum system in which
the wavefunction has a macroscopic quantum effect, due to the
unique condensate of electron pairs. The amplitude of the
wavefunction is directly related to the pair density, but both
amplitude and phase enter the Josephson current\,: the coherent
tunneling of pairs between superconductors. Very sensitive devices
exploit the superconducting state, however properties of the
{\it condensate} on the {\it local scale} are largely unknown, for instance, in
unconventional high-T$_c$ cuprate, multiple gap, and gapless superconductors.


The technique of choice would be Josephson STS, based on Scanning
Tunneling Spectroscopy (STS), where the condensate is {\it
directly} probed by measuring the local Josephson current (JC)
between a superconducting tip and sample. However, Josephson STS
is an experimental challenge since it requires stable
superconducting tips, and tunneling conditions close to atomic
contact. We demonstrate how these difficulties can be overcome and
present the first spatial mapping of the JC on the nanometer
scale. The case of an MgB$_2$ film, subject to a normal magnetic
field, is considered.

\end{abstract}

\section{\bf Introduction}

As Landau first suggested, the superconducting (SC) state is a
quantum condensate represented by a macroscopic wavefunction $\Psi
= \Psi_0 \ e^{i\,\varphi}$, accounting for the two fundamental
properties\,: zero resistance and perfect diamagnetism.
This SC wavefunction varies on the scale of the
coherence length $\xi$\,: when the superconductor is in
contact with a normal metal, or within the vortex core.
However, up to present, local variations of the SC state have not
been measured in a direct way, but rather inferred indirectly.

Local electronic properties are probed by scanning tunneling
microscopy/spectroscopy (STM/STS), where the tunneling current
between an atomically sharp tip and a sample as a function of the
bias voltage, $I(V)$, is measured. Atomic or larger-scale images of
the surface can be done by scanning the tip (topographic mode), or
at a given point, $I(V)$ curves are locally acquired
(spectroscopic mode). Usually the tip is made of a normal metal (W
or Pt/Ir) whose density of states (DOS) near the Fermi level is
roughly constant. Then the differential conductance, $dI/dV$, is
proportional to the sample local DOS (at the energy $E_F + eV$).

An important advance is STS\,: the combination of topographic imaging
with tunneling spectroscopy, resulting in high resolution conductance maps
(equivalently DOS maps). Using a normal metal tip to
study superconductors, STS measures the {\it quasiparticle} (QP)
DOS, first derived by Bardeen, Cooper and Schrieffer, which
generally reveals a gap 2\,$\Delta$ at the Fermi level. Giaever
\cite{Giaever} verified the BCS model using planar junctions in
which a thin oxide layer separates a normal metal electrode from a
superconducting one (SIN junction) or two superconductors (SIS
junction). In their pioneering work, Hess et al.
\cite{Hess1,Hess2} extensively studied the Abrikosov vortex
lattice in 2$H$-NbSe$_2$, in a magnetic field, using low
temperature STS. However, there one observes the changes of the
quasiparticle DOS due to the suppression of the gap in the vortex
cores; it is not a direct measurement of the condensate.


Measuring the SC condensate is particularly needed in the case of
high-T$_c$ cuprates, where the microscopic mechanism is still
unknown. STS conductance mapping has shown that a {\it pseudogap}
at the Fermi level, existing within the vortex core \cite{Renner},
is also induced by disorder
\cite{disorder,Cren_maps,Pan_inhomo,Howald} with a shape similar
to the one found above $T_c$. Thus, the origin of this pseudogap
is a key question \cite{Deutscher} and the quasiparticle DOS
measurement alone cannot decide on its physical origin. The QP
spectrum does not suffice for gapless superconductivity, where the
conductance gap is locally vanishing due to supercurrents or magnetic
impurities, while a condensate exists. A direct probe of the
Cooper pair density is therefore needed.


As predicted by Josephson \cite{Josephson} in 1962, and first
verified by Anderson and Rowell \cite{Anderson}, a distinct
current can flow between two superconductors separated by a thin
layer: the tunneling of Cooper pairs. The Josephson current (JC)
is a supercurrent due to the phase difference between the SC
wavefunctions of the electrodes, allowing a DC flow for zero bias
voltage across the junction. The Josephson effect is not only the
basis of very sensitive and fast switching devices, but it is the
measurable quantity directly connected to the quantum condensate.
Recently two groups \cite{Dynes,Rodrigo} have been able to measure
the JC using low temperature STM, but only at a single point. In
this Letter we report the first mapping of the Josephson current,
using a superconducting MgB$_2$ tip as a novel STS probe.

\section{\bf Towards Josephson spectroscopy}

The first step towards a Josephson STM is to realize stable SC
tips showing characteristic spectra both in the SIN (normal
surface) or SIS (superconductor surface) cases. In the former, the
observed conductance gap is just $2 \Delta$, while in the SIS
configuration, it is $2(\Delta_{tip} + \Delta_{sample})$, see
Fig.\,1. In comparison, the SIS singularities are very sharp, due
to the convolution of the tip and sample DOS. A second set of
peaks, for finite $T$, are characteristic of the SIS junction and
occur at $\pm(\Delta_{tip} - \Delta_{sample})$; see right inset of
Fig.1. Pan et al. \cite{Pan_tips} showed both SIN and SIS
tunneling using a Nb tip, while later Naaman \cite{Dynes} and
Rodrigo \cite{Rodrigo} et al. used Pb tips, or Pb covered in a
thin layer of Ag. Our results in Fig.1 are very similar to these
works.

\begin{figure}[h]
\vbox to 5.6cm{
    \epsfysize=5.8 cm
    \hskip 3.5cm \epsfbox{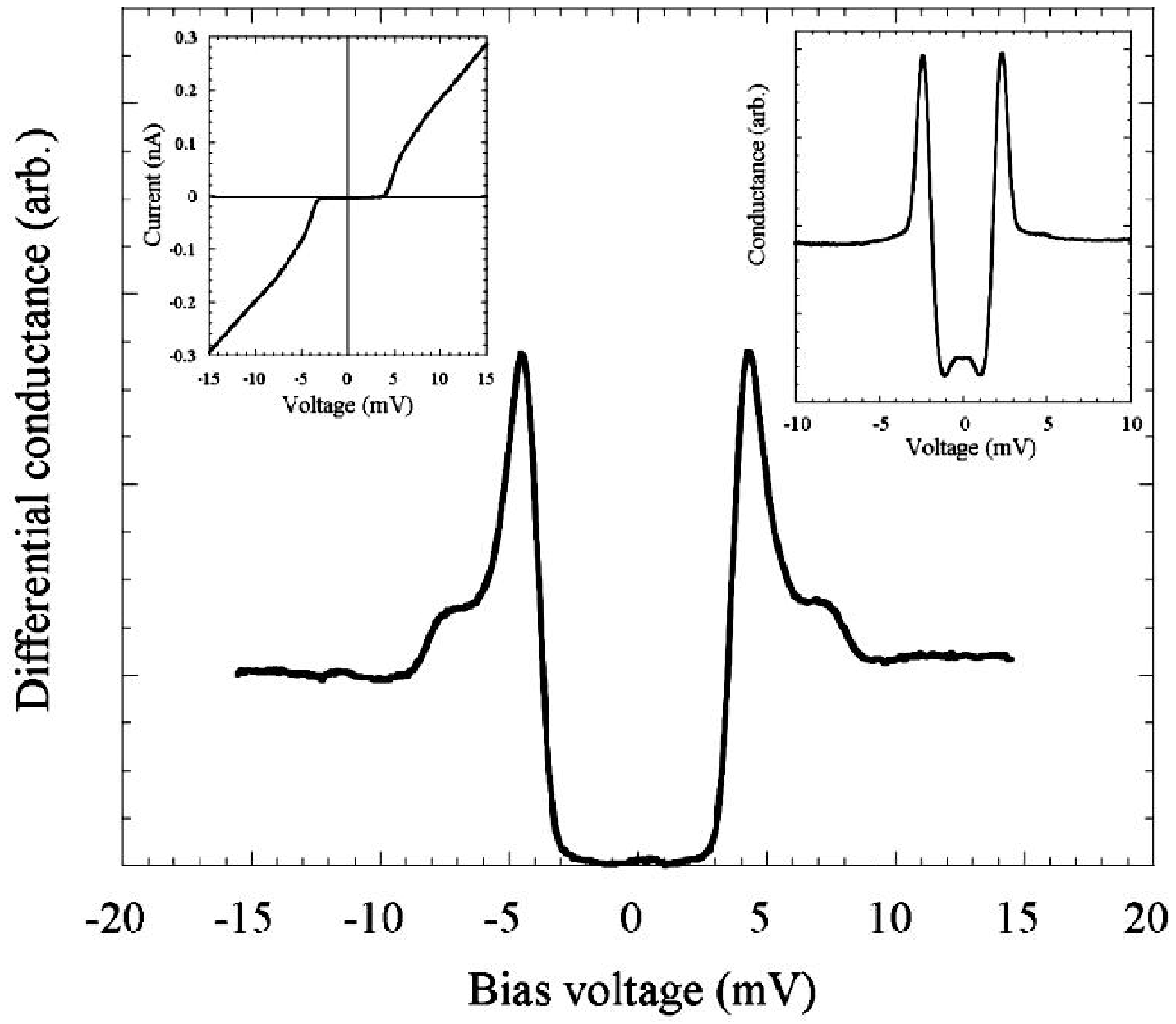}
    }
    \end{figure}
{\vskip 1 mm
    {\small Fig.1. SIS conductance and I-V (Left Inset)
    curves, taken at $T = 2 K$, using a single crystal MgB$_2$ tip
    and an MgB$_2$ sample. Right Inset\,: Conductance spectrum for Nb tip/NbSe$_2$ sample
     ($T=4.2 $\,K).}}

\vskip 2 mm

We produce SC tips with two methods, either by mechanically
breaking a Nb wire in the STM vacuum chamber \cite{Kohen_tips}, or
by gluing a small MgB$_2$ single crystal to a PtIr wire
\cite{Giubileo_gaps}. Both procedures yield tips having
distinctive SC properties, as revealed by SIN and SIS
spectroscopy. MgB$_2$ is a two-band superconductor
\cite{Giubileo_gaps,Souma_mgb2} offering different signatures in
the quasiparticle SIS characteristics, depending on the tip (or
sample) surface. If the sample is an MgB$_2$ thin film, as in
Fig.1, with {\bf c} axis normal to the film, the small gap
$\Delta_\pi$ dominates the quasiparticle current\,: the total
conductance gap seen in Fig.\,1 is then $4 \Delta_\pi \simeq 10$
meV as expected \cite{Bobba_film}. The large gap, $\Delta_\sigma
\simeq 7$ meV, gives secondary peaks whose resolution depends on
the local tunneling geometry. This is a particular electronic
structure effect, since the tunneling into the 3D $\pi$ band has a
higher probability than to the 2D $\sigma$ band.

\begin{figure}[h]
\vbox to 5.2cm{
    \epsfysize=5.6 cm
    \hskip 5cm \epsfbox{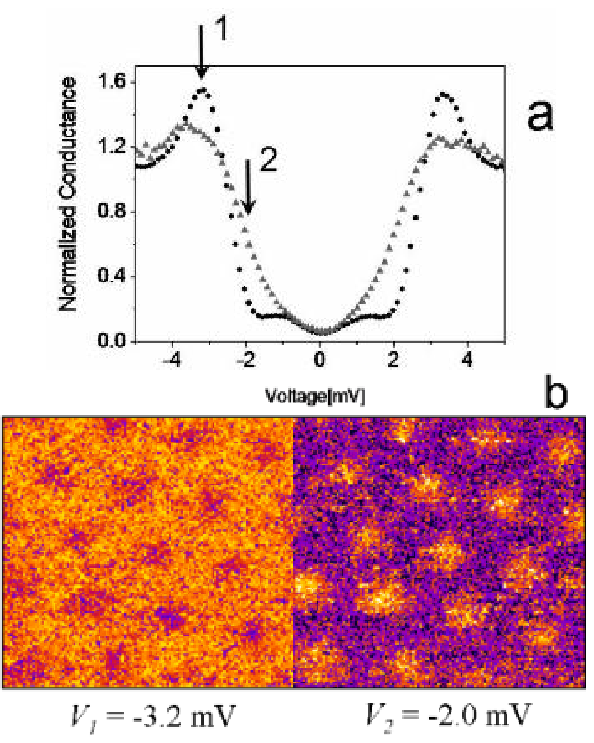}
    }
    \end{figure}
{
    {\small Fig.2. STS scan of the vortices in 2{\it H}-NbSe$_2$ using a superconducting MgB$_2$
    tip. (a) Normalized conductance measured in the vortex core (SIN), triangles, and
    in between the vortices (SIS), dots. \\ (b) Raw conductance maps at the
    indicated voltages (1 and 2 in {\bf a}). Scan range\,: 380 nm, magnetic field\,: 0.25 Tesla,
    Temperature\,: 4.2 K.
     }}
\vskip 2 mm

The second step is to realize STS mapping in the SIS
configuration. While atomic resolution in the topographic mode has
been previously obtained \cite{Kohen_tips,Giubileo_gaps,Pan_tips},
to our knowledge {\it scanning} spectroscopy has not been achieved
with SC tips. However, it is a basic requirement for the Josephson
microscopy. Probing the vortex lattice in 2{\it H}-NbSe$_2$, using
either MgB$_2$ or niobium tips, tests the spatial resolution of
the sample DOS and the stability of the tip SC properties in the
magnetic field. In Fig.2{\bf b} we show the vortices, obtained with
an MgB$_2$ tip, clearly revealed in the QP conductance maps at the
selected bias voltages\,: $V_1$ at the QP peak and $V_2$ within
the conductance gap. The contrast reversal of the maps, for these
two voltages, is expected\,: between the vortices the conductance
is of the SIS type, while in the vortex core the junction becomes
SIN (see Fig.2{\bf a}). A complete discussion of the
bias-dependant conductance maps, in the SIS geometry with a Nb
tip, can be found in \cite{proslier}.

The penultimate step towards Josephson microscopy is to measure
the JC systematically throughout the sample. In this regard, the
JC measurement is quite different from conventional STS since the
probability for pair tunneling is much smaller than for single
quasiparticles \cite{Smakov}. Consequently the tunneling
resistance must be set much lower, e.g. 50 k$\Omega$, as compared
to the usual $\sim$ 100 M$\Omega$, maintained high so that the tip
never touches the sample. As given by Ambegaokar and Baratoff
\cite{Baratoff}, assuming identical superconductors at $T = 0$,
the $I_c R_n$ product is\,: $I_c R_n = \pi \Delta/2\,e$ where
$R_n$ is the normal (QP) resistance. The Josephson coupling
energy, $E_J = (\hbar/2e) I_c$ must be larger than $k T$, or else
thermal smearing prevents the effect. Then one has the condition:
$$
R_n < \frac{\Delta}{k T} \ R_0
$$
where $R_0$ is the quantum resistance ($\hbar/e^2$). As pointed
out by Smakov et al.\cite{Smakov}, with usual values of the gap
and $T$, it is quite difficult to measure the JC without requiring
direct contact between tip and sample\,: $R_n$ is close to $R_0$.
In the remainder of this work we use MgB$_2$ tips, where the JC is
stronger than for Nb, for which one estimates, at $T = 2$ K, $R_n
< 60\,$k$\Omega$.

Naaman, Teizer and Dynes \cite{Dynes} have measured the JC, using
Pb/Ag tips on a Pb sample, in the case where the above inequality
is only roughly verified\,: the result is fluctuation-dominated
pair tunneling. Indeed, Ivanchenko et Zil'berman \cite {Ivan} have
shown that the JC for very small junctions is strongly affected by
the fluctuations of the relative phase $\Delta \varphi$ between
the SC condensates of the two electrodes. For a voltage biased
junction, a characteristic $dI(V)/dV$ curve is obtained, with a
sharp peak at zero bias. Our JC measurements using the MgB$_2$
tip, Fig 3, reveal the same qualitative $I(V)$ and conductance
curves as those of Ref. \cite{Dynes}. The spectra were obtained in
succession by lowering the normal tunneling resistance $R_n$ from
31 down to 14 k$\Omega$, with a final $I_c R_n$ product of 4 meV.
Raising the resistance again gives back a similar set of $I(V)$
curves. Based on the theory of \cite{Ivan}, and the correct
response of the junctions subject to a micro-wave field by Naaman
et al. \cite{Dynes}, this JC is directly related to the pair
density of the sample.

\begin{figure}[h]
    \vbox to 5.2 cm{
    \epsfxsize=5.6cm
    \hskip 4cm \epsfbox{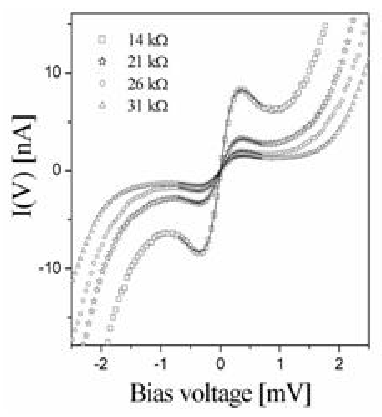}
    }
    \end{figure}

{\vskip 1 mm
    \small Fig.3. $I(V)$ spectra for the MgB$_2$-MgB$_2$ junction
    (STM at a single point, $T=2.1 K$).
    Inset\,: corresponding tunneling resistance values ($R_n$).
    Solid line\,: fit using the theory of \cite{Ivan}.
     }
\vskip 2 mm

The solid lines in Fig.\,3 are the fits using the theory of Ref.
\cite {Ivan} in which the phase fluctuations are due to the
voltage noise at the junction, represented by an `effective
temperature' $T_{eff}$\,: $<v(t)v(0)>=2\,R\,k\,T_{eff} \delta(t)$,
where $R$ is the source resistance. As part of this noise
originates from the STM electronics at 300 K, $T_{eff}$ can be
much higher than the actual temperature of the junction (2.1 K).
Using $\Delta=2.5$ meV from the SIS spectra, and knowing $R_n$, we
deduce from the series of fits : $T_{eff} \simeq 66$ K and $R
\simeq$  100 $\Omega$, close to the values of \cite{Dynes} for a
similar electronics\,: $T_{eff} = 56$ K and $R \simeq 80\,
\Omega$. This fluctuating voltage $\sim 50\, \mu$V does not affect
the QP spectrum, as in Fig.\,1, while it strongly does in Fig.\,3
(JC) through the phase fluctuations. For superconductors having a
small gap compared to MgB$_2$, in addition to lower temperature,
it may be necessary to further reduce the high-frequency voltage
noise reaching the junction. Besides filtering, part of the
electronics can be placed at low temperature, on or near the STM.
In our case, Fig.\,3, the agreement between theory and experiment
is satisfactory\,: the conductance peak measured at zero bias is
indeed the fluctuating JC.

\section{\bf Josephson conductance mapping}

The final step is to perform scanning while measuring the JC in
the STM spectroscopic mode. However, a high resistance is needed
during the tip scan, to avoid irreversible tip damage, but a low
resistance is needed to measure the JC. We have redesigned our STS
acquisition \cite{Cren_maps} (i.e. a complete I-V spectrum at each
pixel of an image) and have developed a new `sewing needle'
scanning mode. In this mode, the tip is approached towards the
surface (lower $V$, higher $I$) in order to obtain a much smaller
resistance (in the 60 kOhm range, or less). An I-V spectrum is
acquired. The tip is then retracted from the surface (higher $V$,
lower $I$), moved to the next point of the image, and the cycle is
repeated. This `sewing needle' mode is an essential ingredient of
the Josephson STM, where the feedback remains active when lowering
and stabilizing the tip prior to acquiring the spectrum. In this
way, the complete STS data set consists of 256 JC conductance maps,
$dI/dV(V,x,y)$, measured simultaneously with the topography.
A similar data set is acquired for the QP conductance maps.

We now discuss how the spatial maps of the JC correlate to the QP
conductance maps and to the topographic map. Here we consider an
MgB$_2$ thin film, with {\bf c} axis oriented normal to the
substrate and subject to a magnetic field of 0.18 Tesla, applied
normal to the surface. In Fig.4, the principal results of a
complete STS scan over a selected region of the film are
summarized. Both Josephson and quasiparticle DOS maps were
acquired in this {\it identical} region\,: 256 images are
simultaneously acquired where in the first, the bias voltage range
is from -2 to +2 meV, and in the second, from -15 to +15 meV.

The topography is shown in image {\bf a} revealing six or so
different grains. In comparison, the quasiparticle DOS maps are
complex and rich in information, with variations of both the
conductance gap width and the peak intensity. To simplify the
discussion, we display in {\bf c} a color map of the quasiparticle
peak intensity at the selected voltage $V=- 5$ mV, corresponding
to the nominal SIS peak of the MgB$_2$-MgB$_2$ junction (i.e. e$V$
= 2$\Delta_\pi$). In such a way, a small or smoothed gap appears
dark, and a sharp conductance peak appears bright. Analogously, we
display in map {\bf d} the JC conductance peak intensity at zero
bias (see {\bf f}). While the two maps {\bf c} and {\bf d} have a
different origin (i.e. QP conductance in {\bf c} and JC in {\bf
d}), they show a definite spatial correlation between them.

\begin{figure}[h]
    \vbox to 4.6 cm{
    \epsfxsize=6.4cm
    \hskip 3.5 cm \epsfbox{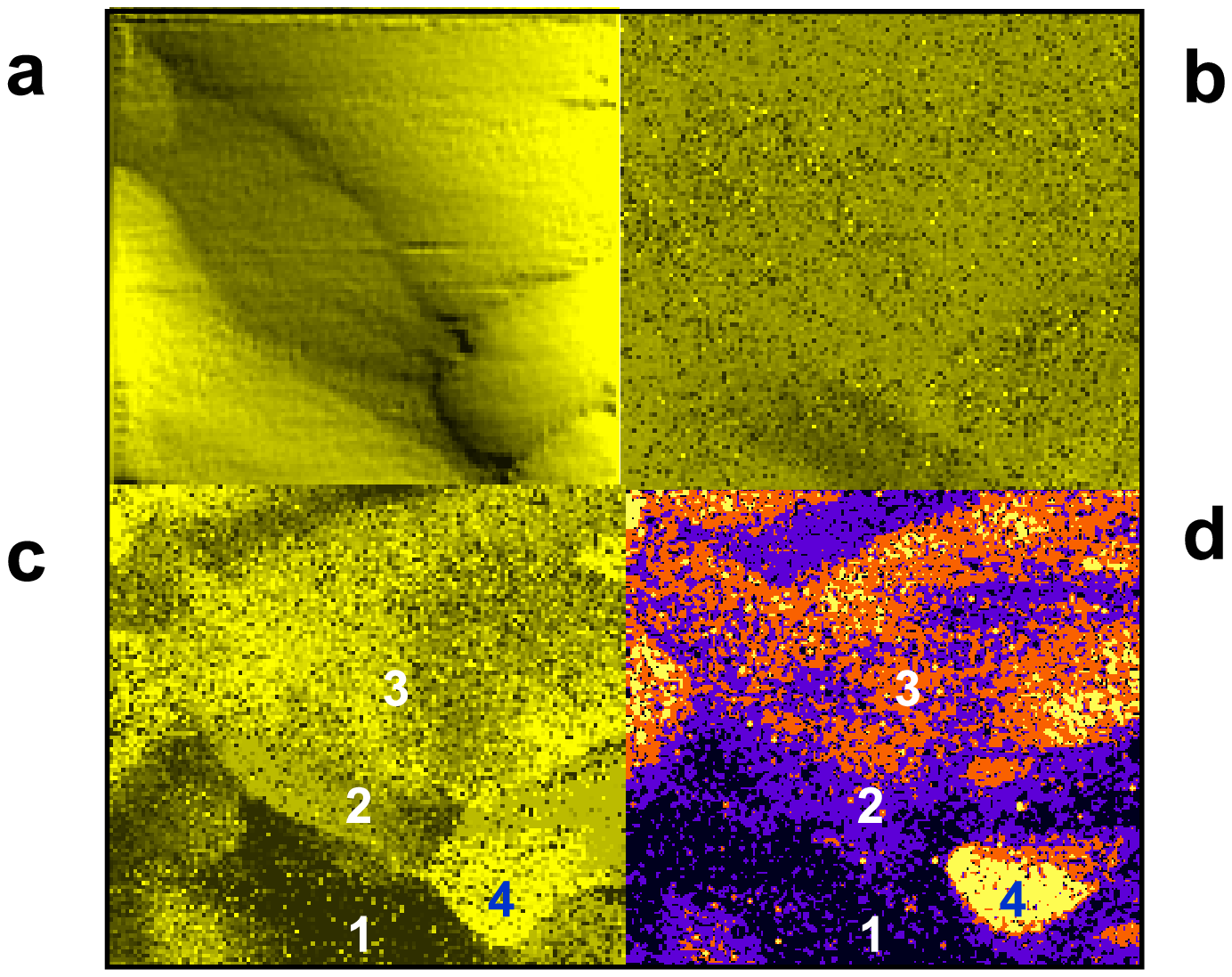}
    }
\end{figure}

\begin{figure}[h]
    \vbox to 4.8 cm{
    \epsfxsize=7.6 cm
    \hskip 3cm \epsfbox{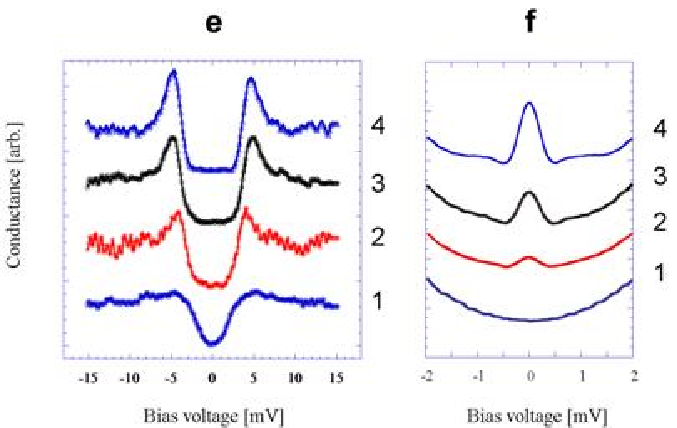}
    }
\end{figure}
{
    \small Fig.4. Analysis of a complete STS scan of a 220 nm x 220 nm area of
    the MgB$_2$ thin film, and using the MgB$_2$ tip,
    representing a total of 2$\times$128$\times$128 $I(V)$ spectra\,: \\
    {\bf a}. The topography (standard STM Z-deflection),\\
    {\bf b}. Josephson conductance map at $V$ = 2 mV\\
    {\bf c}. Quasiparticle peak intensity map, at $V$ = -5 mV,\\
    {\bf d}. Josephson conductance map, at $V$ = 0 mV\\
    Regions 1-4, defined by the colour scale of image {\bf d}, reveal the characteristic spectra\,: \\
    {\bf e}.\ Quasiparticle conductance spectra (regions 1-4),\\
    {\bf f}.\ Josephson conductance spectra (regions 1-4)\\
     \null \ \ \ In {\bf e} and {\bf f} the curves are normalized
     to the same value, then offset for clarity.}
\vskip 2 mm

The entire scan area (220 nm x 220 nm) can consequently be divided
into regions where the SC state has similar characteristics.
Furthermore, by comparison to the topographic image ({\bf a}),
some of the regions can be identified with individual grains. This
is not always the case, there are two dark regions, at the lower
part and the very upper part of the QP image ({\bf c}), where the
superconducting state is either very weak (upper dark region) or
totally absent (lower dark region). In this experiment, as the
MgB$_2$ film was subject to a normal field, magnetic flux is
present at the locations of strong defects, but no vortex lattice
is observed. Furthermore, as will be shown, the spectra acquired
within these regions lead to the conclusion that they correspond
to a vanishing order parameter.

The complete STS data set allows to select precisely the spectra
corresponding to the regions 1-4, as defined in the figure, for
direct comparison. The QP conductance and the JC spectra are
plotted in Fig.4{\bf e} and Fig.4{\bf f}, respectively. It is
immediately evident that the general trend of the JC curves
follows the trend of the QP curves. The strongest JC peak occurs
in region 4, precisely where the SC gap has the most significant
SIS characteristics (strongest peaks and deep gap). On the
contrary, in region 1, the Josephson peak is totally absent and
the QP spectrum is typical of the SIN type, i.e. the pair
potential ($\Delta$) is locally vanishing. Regions 2 and 3 are
thus intermediate, and they display a less-pronounced Josephson
peak, accompanied by weaker SIS characteristics, with a smaller
conductance gap. For regions 1-4, the local JC intensity
correlates perfectly to the local SIS spectra.


The Josephson conductance map at the selected value of $V$=2.0
meV, image {\bf b}, is shown for comparison to the JC map at zero
bias, {\bf d}. Aside from the `normal' region 1 discussed above,
there is practically no change in the spatial characteristics (it
appears uniform). This is an important experimental verification
of the STS\,: at this particular voltage ($V$=2.0 meV), beyond the
Josephson peak and at the same time well below the SIS peak, there
is no significant spatial variation of the conductance spectra.

\begin{figure}[h]
    \vbox to 5.4cm{
    \epsfxsize=6.2 cm
    \hskip 3 cm \epsfbox{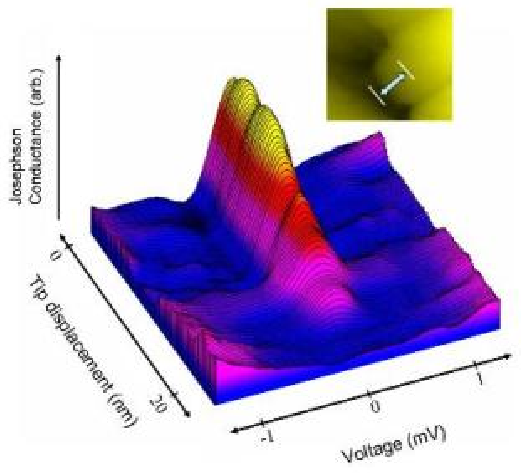}
        }
    \end{figure}
{
    \small Fig.5. Three-dimensional view of the change in the
    Josephson conductance peak, as a function of the tip position,
    along the line (Inset) from region 1 to the center of region 4.
    The plot is from the identical STS data set of Fig.4.
    Thus the Josephson current increases from nearly zero to its
    maximum intensity in a distance of about 15 nm.
    Inset\,: Zoom of the topography (from Fig.4{\bf a}). \\
    }
\vskip .5 mm

Region 4 is of particular interest in that first, it has the
highest Josephson conductance peak and second, it correlates to a
particular grain observed in the topography. Here the $I_c R_n$
product is about 4 meV, at its maximum value. We note the
variation of the JC as one crosses from the `normal' region 1 to
region 4 across the grain boundary\,: the change is continuous,
but quite abrupt, as shown as a 3D plot in Fig.5. The approximate
value for the distance over which the JC peak evolves is $\sim$ 15
nm. Note that the QP conductance follows the shapes 1-4, as in
Fig.4{\bf e}.

The JC peak also decays in the `normal' region in the upper part
of Figs.\,4 {\bf c} or {\bf d}, where no particular topographical
structure is present. Crossing the entire region along a line, the
variation of the JC and the QP conductance shapes are compared in
a top-view in Fig.6. Clearly, the JC peak vanishes, on the scale
of $\sim 25$ nm, and reappears again. The parallel effect on the
QP spectra, while following the same trend, is less pronounced.

The physical effect is clearly due to the presence of magnetic
flux in this region. The evolution of the SIS peaks (right panel
of Fig.6) is first the lowering of the peaks due to the
supercurrents (Doppler shift) and second, the vanishing of the
pair potential, i.e. the junction becomes SIN. We have recently
found this precise behavior probing a single vortex with a SC
niobium tip\cite{proslier}. Quantitatively, the coherence length
as inferred from the vanishing of the JC peak in Fig.6, $\xi \sim
25$ nm, is smaller than the value of Eskildsen et al.
\cite{Fischer} ($\xi \sim$ 57 nm) from the DOS crossing the $\pi$
band vortex core, for a single crystal. As MgB$_2$ is a two-band
superconductor, with a significant coupling between the bands, the
effective coherence length depends on the particular experiment
and the type of sample (here, a thin film). A lower bound, $\sim$
13 nm, is estimated from the second critical field, $H_{c2}$,
while the weak superconductivity of the $\pi$ band gives an upper
value of $\sim$ 60 nm.

\begin{figure}[h]
    \vbox to 4.6cm{
    \epsfxsize=6.0 cm
    \hskip 3cm \epsfbox{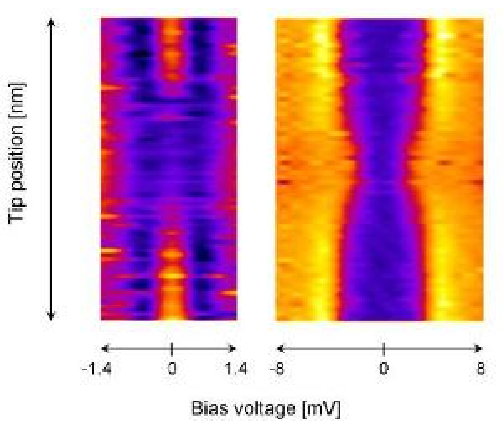}
    }
    \end{figure}
{
    \small Fig.6. Top-view of the STS scan across
    the `normal' zone (upper part of Fig.\,4{\bf c}) to compare the JC (left)
    to the QP (right) variations.
    }
\vskip 2 mm

There are also more intricate variations in the Josephson map
({4\bf d}) as seen in the boundaries of regions 2 and 3. The
presence of grain boundaries or impurities may affect the local
electronic properties, in addition to the magnetic flux. Due to
the coupling between the bands, there could be a contribution to
the JC from the Cooper pairs from the $\sigma$ band, changing with
the (local) tunneling junction. Work is in progress to evaluate
this effect but, in a first approximation, $I_c R_n$ is
proportional to the small gap, $\Delta_\pi$.

The main goal of STS mapping of the JC is therefore achieved. The
spatial evolution of the JC is in qualitative agreement with the
quasiparticle DOS maps (Fig.4), in a first approximation, which is
expected for conventional SC. For the first time the suppression
of the SC state, due to the applied magnetic flux, is observed in
the JC directly (Fig.6). There is a long list of possible
applications of such a probe. It could be used to map the
condensate of gapless superconductors, different vortex states,
and be of general use in the case of structural changes, such as
steps or boundaries, and point defects. It could test the $d$-wave
symmetry of the high-Tc superconductors, with or without these
perturbations. Clearly, the origin of the pseudogap is a key
question where the observed gap, inferred from the DOS, is not
directly the order parameter, and a second energy scale accounts
for the transition \cite{Deutscher}. A direct probe of the Cooper
pair density, such as with the Josephson STS, may help answer the
question.

{\small We gratefully thank F. Breton and F. Debontridder for
their technical assistance. This work was supported by the project
GBP {\it Mat\'{e}riaux aux propri\'{e}t\'{e}s remarquables}}.

\end{document}